\documentclass[11pt]{article}

\renewcommand{\epsilon}{\varepsilon}

\usepackage[figuresright]{rotating}

\usepackage{bm}
\usepackage{bbm}
\usepackage{amsfonts, amsmath}
\usepackage{enumitem}
\usepackage{xcolor}
\usepackage{booktabs}
\usepackage{natbib}
\usepackage[margin=1in]{geometry}
\usepackage{graphicx}
\usepackage{setspace} 

\newcommand{\x}{\bm{x}}

\newcommand{\ncancer}{K}
\newcommand{\ub}{\bm{u}}
\newcommand{\X}{\bm{X}}
\newcommand{\Ub}{\bm{U}}
\newcommand{\one}{\mathbbm{1}}

\newcommand{\alphab}{\bm{\alpha}}
\newcommand{\betab}{\bm{\beta}}
\newcommand{\omegab}{\bm{\omega}}

\newcommand{\R}{\mathbb{R}}
\newcommand\numbereqn{\addtocounter{equation}{1}\tag{\theequation}}
\DeclareMathOperator*{\argmax}{arg\,max}

\usepackage{lineno}

\newtheorem{remark}{Remark}[section]

\allowdisplaybreaks
\title{Using the ``Hidden'' Genome to Improve Classification of Cancer Types}

\author{
	Saptarshi Chakraborty, 
	Colin B. Begg,
	and \\ 
	Ronglai Shen (\texttt{shenr@mskcc.org}) \\ \\
	Department of Epidemiology \& Biostatistics,\\ 
	Memorial Sloan-Kettering Cancer Center \\ New York, NY 10017, USA
}

\date{}

\begin{document}

	\maketitle

	\label{firstpage}
	
	
		\begin{abstract}
		\emph{It is increasingly common clinically for cancer specimens to be examined using techniques that identify somatic mutations. In principle these mutational profiles can be used to diagnose the tissue of origin, a critical task for the 3-5\% of tumors that have an unknown primary site. Diagnosis of primary site is also critical for screening tests that employ circulating DNA. However, most mutations observed in any new tumor are very rarely occurring mutations, and indeed the preponderance of these may never have been observed in any previous recorded tumor. To create a viable diagnostic tool we need to harness the information content in this “hidden genome” of variants for which no direct information is available. To accomplish this we propose a multi-level meta-feature regression to extract the critical information from rare variants in the training data in a way that permits us to also extract diagnostic information from any previously unobserved variants in the new tumor sample. A scalable implementation of the model is obtained by combining a high-dimensional feature screening approach with a group-lasso penalized maximum likelihood approach based on an equivalent mixed-effect representation of the multilevel model. We apply the method to the Cancer Genome Atlas whole-exome sequencing data set including 3702 tumor samples across 7 common cancer sites. Results show that our multi-level approach can harness substantial diagnostic information from the hidden genome.}
	\end{abstract}


		\section{Introduction}
	\label{sec:intro}
	
	Identifying the anatomic site of origin of malignancy is a crucial problem in cancer diagnosis, including cancers of unknown primary which make up 3-5\% of total cancer diagnoses worldwide and are associated with poor prognosis \citep{varadhachary:raber:2014, conway:2018}, as well as in the more recent context of ``liquid biopsies'' whereby circulating tumor DNA (ctDNA) is identified in the blood and can in principle be used for early detection of cancer \citep{diaz:bardelli:2014, wan:2017}. In recent years, large-scale sequencing studies have begun to reveal the somatic mutational landscape in cancer. Evidence emerging from these studies suggests that common somatic mutations can be highly organ site specific \citep{haigis:2019}. This serves as a  basis for identifying site of origin using the mutational profile identified from the tumor sample or from ctDNA sequencing.
	
	Predicting cancer types based on mutations, however, comes with its own sets of challenges.  Advances in modern genome sequencing technologies and their application to large cancer patient cohorts have resulted in the discovery of millions of unique somatic mutations (variants) in the tumor genome. It is well known that a relatively small number of variants appear in tumors frequently, but the vast majority of variants occur extremely infrequently. Data from large-scale sequencing studies such as the Cancer Genome Atlas \citep[TCGA,][]{Bailey:2018} revealed that over 90\% of the somatic variants in the study were singletons, i.e. observed only once in the $>$10,000 sequenced tumor samples. The vast trove of these extremely rare variants, routinely ignored or under-utilized, may in fact  harbor important signals that can be harnessed for clinically relevant problems. A statistical and computational solution to extract relevant information from these rare variants is needed. 
	
	Furthermore, there are many more variants yet to be discovered, and virtually every gene sequencing experiment identifies a multitude of previously unseen mutations. This makes variant-based prediction of cancer types extremely difficult: in sequencing of a new tumor, for many (if not all) mutations in the tumor there will be little or no direct previous information.  One approach is to perform classification only at the gene level \citep{soh:2017}. That is, instead of using the individual variants directly, the genes in which the mutations occurred are used as predictors. This strategy aggregates all (including rare) variants in a gene and reduces the total dimension of predictors, as the total number of genes is much smaller than the total number of possible variants. Furthermore, all genes are known, and hence the problem of encountering predictors that were not present in the training data does not arise. However, this approach ignores variant-level information that can be critical. For example, common variants in the \textit{KRAS} gene are known to be highly tissue-specific: the \textit{KRAS} G12C variant is primarily associated with lung adenocarcinoma, while \textit{KRAS} G12R occurs more exclusively in pancreatic cancers. Thus, a gene level classification model is likely to produce sub-optimal results as such variant-level discriminant information is lost. 
	
	
	We note that the problem of tumor classification based on mutation profiles has analogies with content-based text classification problems in natural language processing. The work of \cite{taskar:wong:koller:2003}  is particularly noteworthy in this context.  The authors suggest ``learning" the role of a previously unseen word (feature) through its context, i.e., neighboring words/features, while classifying a new text document. In the field of machine learning in general the problem of classification/prediction with previously unseen features can be framed as a problem of domain adaptation or transfer learning  \citep{shi:knoblock:2017}, where a model trained on a specific source domain is adapted to a related but different target domain. Existing approaches to domain adaptation problems include tree based ensemble methods \citep{habrard:2013}, nearest neighbor regression/classification techniques \citep{shi:knoblock:2017}, and neural networks \citep{ganin:2016:domain}. However, as is often the case in the machine learning literature, many of these approaches are black-boxes, and do not aid interpretable statistical inference on the model parameters (and thus on the predictors).

	The major contribution of this paper lies in the development of a rigorous statistical  framework that through the associated DNA sequence context infers the effects of previously unseen variants and subsequently uses these inferred effects in the prediction of cancer sites. More precisely, we build a multi-level classification model with an embedded hierarchical meta-regression step in which the  effects of \textit{all} (including previously unseen) variants are modeled as functions of a number of known \textit{meta-features} describing the associated mutation contexts. Through the meta-regression, these meta-features can be used to quantify the effects of previously unseen variants in a new tumor, and simultaneously provide a  highly informative yet substantially smaller dimensional projection of the ultra high-dimensional (in the order of millions) predictor space.  In our example we employ two categorical meta-features, the gene itself and the single base substitution membership of the variant, though the model can easily be extended to include other meta-features.

	 Through an equivalent mixed-effect representation of our multi-level model, we demonstrate how these meta-features affect classification. In particular we note how our approach  rigorously combines information from the raw variants as well as from  the genes in which the variants originate, thus providing a refinement and extension to the existing gene-level approaches \citep{soh:2017}. Careful assignment of a shrinkage prior as a further hierarchical layer in the mixed effect model aids a group-lasso regularized likelihood of the model parameters of interest, thus enabling straight-forward implementation. The parameter estimation is performed following a mutual information-based feature screening \citep{cover:thomas:2005}. This makes our approach highly scalable, allowing incorporation of possibly millions of variants and appropriately chosen meta-features. We apply the method to the  publicly available TCGA mutation data set, electing to focus on 7 common cancer types.
	
	
	The remainder of the article is organized as follows. In Section~\ref{sec:tcga_data_descr}, we provide a brief description of the TCGA data. In Section~\ref{sec:model} we first provide some intuition on our variant and meta-information based approach, then formalize the intuition in terms of a (sparse) multi-level multinomial logistic classification model, and finally describe an implementation strategy for our model that can handle millions of variants as predictors. Section~\ref{sec:application_tcga} describes the application of our model to the TCGA data set. Finally, in Section~\ref{sec:discuss} we address strengths and  limitations of our method, and research directions that will be pursued in future. Additional technical and numerical details are included in the Appendix.

	\section{Data source and description} \label{sec:tcga_data_descr}
    The somatic variant data are  from the Cancer Genome Atlas (TCGA) study derived from whole-exome sequencing of approximately 10,000 tumor samples across multiple cancer types \citep{Bailey:2018}. We focus our attention on the 7 most prevalent tumor types (as determined by the SEER \citep{duggan2016surveillance} program of the National Cancer Institute), namely  bladder urothelial carcinoma (BLCA), breast invasive carcinoma (BRCA),  colorectal adenocarcinoma (COADREAD), lung adenocarcinoma (LUAD), pancreatic adenocarcinoma (PAAD), prostate adenocarcinoma (PRAD),  and  skin cutaneous melanoma (SKCM). These have sample sizes (number of tumors) of 411, 1025, 559, 568, 176, 495, and 468 respectively.  Among these 3702 tumors a total of 811,497 unique somatic variants were identified, the majority of which were rare variants (see Table~\ref{tab:rNr}). Specifically, 95\% of all observed variants are singletons (appearing only once in the cohort).
    	
    \begin{table}[ht]
		\setlength{\tabcolsep}{8pt} 
		\renewcommand{\arraystretch}{1.25} 
		\centering
		\label{tab:rNr}
		\begin{tabular}{|r|rrrrrrrrrr|}
		  \toprule
            $r$ & 1 & 2 & 3 & 4 & 5 & 6 & 7 & 8 & 9 & 10 \\ 
            \midrule
            0+ & 770057 & 35027 & 4541 & 1077 & 384 & 175 & 77 & 40 & 28 & 19 \\ 
            10+ & 11 & 7 & 4 & 2 & 4 & 3 & 5 & 2 & 1 & 2 \\ 
            20+ & 2 & 1 & 4 & 1 & 0 & 0 & 0 & 0 & 2 & 0 \\ 
            30+ & 1 & 2 & 1 & 2 & 1 & 0 & 1 & 0 & 0 & 1 \\ 
            40+ & 0 & 0 & 0 & 0 & 0 & 1 & 0 & 1 & 0 & 1 \\ 
            50+ & 0 & 0 & 0 & 0 & 0 & 0 & 0 & 0 & 0 & 0 \\ 
            60+ & 0 & 1 & 0 & 0 & 0 & 0 & 0 & 0 & 0 & 0 \\ 
            $\vdots$ & $\vdots$ & $\vdots$ & $\vdots$ & $\vdots$ & $\vdots$ & $\vdots$ & $\vdots$ & $\vdots$ & $\vdots$ & $\vdots$ \\
            260+ & 0 & 0 & 1 & 0 & 0 & 0 & 0 & 0 & 0 & 0 \\ 
            \bottomrule
		\end{tabular}
			\caption{Out of the 811,497 unique variants observed in the 3702 TCGA tumors with one of the seven cancer types, the number of variants $N_r$ that appear exactly $r$ times (i.e., in $r$ tumors), $r = 1, 2, \dots, 270$, are displayed for various $r$'s. The $(i, j)$-th cell shows the observed value of $N_r$ corresponding to $r = 10 (i-1)$ (= left row border) $+ j$ (= top column border). For example, the $(1, 1)$-th entry, 770057, is the observed value of $N_{0 + 1} = N_1$, the number of singletons. The $(4, 3)$-th entry, $1$, tells us that there was a single variant observed 33 times.}
	\end{table}
	
    A key challenge involves making use of new variants emerging in a future tumor sample that have not been observed in the training cohort. In earlier work we showed that 60\% of mutations identified in new tumors are expected to be variants not observed in the TCGA cohort \citep{chakraborty:begg:shen:2019}.

    \section{Multi-level model for cancer type prediction} 	\label{sec:model}
	
    \subsection{Notation and intuition} \label{sec:model_intuition}

\begin{figure}
	    \centering
	    \includegraphics[width= \textwidth]{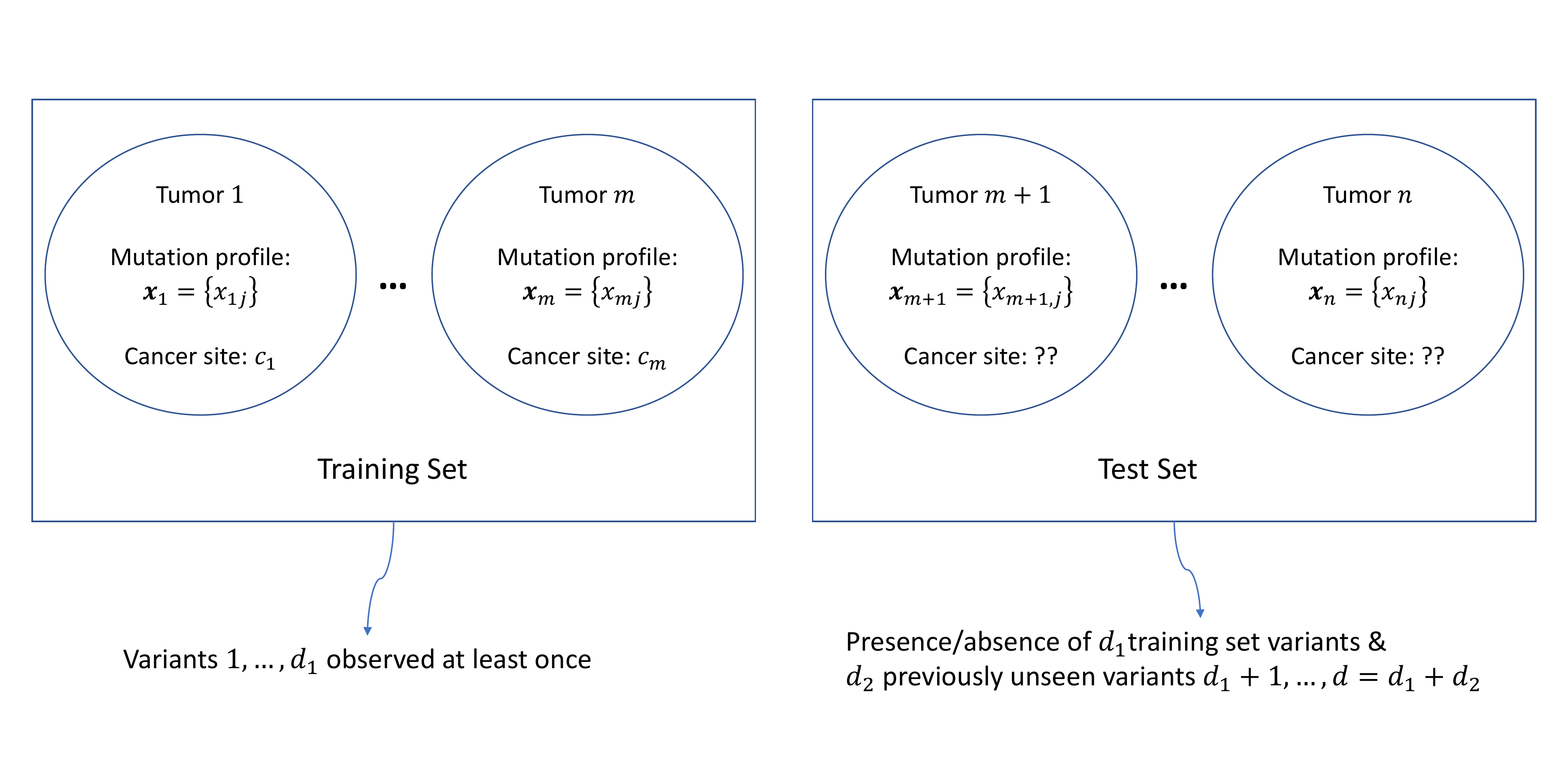}
	    \caption{Diagrammatic representation of the mutation profile based tumor site classification problem.}
	    \label{fig:model_diagram}
	\end{figure}

	Consider a training data set consisting of mutation profiles of $m$ tumors with known cancer types and a test data set of $n-m$ ($n > m$) tumors whose cancer types we aim to predict. For the $i$-th tumor the associated true cancer label is $c_i \in \{1, \dots, \ncancer\}$, where $\ncancer$ is the number of candidate cancer types. We call variants that have been observed at least once in the training cohort as ``recorded" variants, and  we have information on the presence/absence of $d_1$ such recorded variants in the training set. Let $x_{i j}$ denote the binary indicator of the presence/absence of the $j$-th variant in $i$-th individual (0 = absence, 1 = presence), $j = 1, \dots, d_1$. In the test set, we seek to predict the cancer labels based on information on the presence/absence of  $d = d_1 + d_2$ variants in total, where $d_2$ denotes the number of  variants in the test data that do not occur in the training data. A diagrammatic representation of the problem is provided in Figure~\ref{fig:model_diagram}.
	
	We consider a multinomial logistic (``softmax'') regression framework for this classification problem, where for $k = 1, \dots, \ncancer$, the probability $P(c_i = k)$ of the $i$-th tumor having cancer type $k$ is modeled as the softmax (normalized exponential) of linear functions of the predictors $x_{ij}$'s (see Section~\ref{sec:multi_model}). A recorded-variant-only approach fits a classification model to the training data using the raw variants observed in that data set as predictors, and predicts cancer types of test set mutations based on that fitted model. Clearly, this approach can only utilize the information provided by the training set recorded variants, and fails to make use for classification purposes of any information in the multitude of mutations observed only in the test data set. 

	We have developed a multi-level meta-feature regression framework to properly employ of the information in those new (observed only in the test set) variants while classifying  the cancer type of a new tumor.  To this end, we first make a key observation: irrespective of whether a variant is ``recorded'' or ``new'', we always have some quantifiable information obtained from the associated DNA sequencing contexts, called meta-features in this paper. The key idea then involves inferring the effects of new variants from those of the recorded variants with similar meta-features. In terms of a multinomial logistic model, this amounts to connecting the logistic regression coefficients for variants to these meta-features through an appropriate hierarchical meta-regression step. Using this hierarchical meta-regression, the regression coefficients for the new variants in a tumor outside the training set can then be ``predicted'', and subsequently used for classification. The following section formalizes this idea via an appropriate multilevel model.

	\subsection{Model} \label{sec:multi_model}
    Let us denote by $\ub_j = (u_{j, 1}, \dots, u_{j, p})^T$ the values of $p$ meta-features associated with the variants, $j = 1, \dots, d = d_1 + d_2$. Recall that only the first $d_1$ variants are observed in the training data, i.e., the first $m$ of the $n$ total tumors. In this study we include two categorical meta-features on each variant:  the nucleotide substitution signature associated with each variant, and gene membership, and we dummy code each category. More precisely, we define $p$ binary meta-features of which the first 96 indicate whether a variant corresponds to one of 96 possible single base substitution signature categories (see Section~\ref{sec:application_tcga_cv} for more details), and the remaining meta-features are indicators of $\sim 600$ OncoKB cancer genes \citep{chakravarty:2017:oncokb}. We define the design matrix $\X \in \R^{n \times d}$ as $\X^T = (\x_1, \dots, \x_n)$, where $\x_i = (x_{i 1}, \dots, x_{i d})^T$, $i = 1, \dots, n$. (Note that while training we only have access to $x_{i 1}, \dots, x_{i d_1}$, for $i = 1, \dots, m$; the regression parameters associated with the new variants $d_1 + 1, \dots, d$ are not identifiable in the training data set, and hence not estimated.) Also, let $\Ub \in \R^{d \times p}$ be the meta-design matrix defined as $\Ub^T = (\ub_1, \dots, \ub_d)$. This unified representation with all  tumors (training and test combined) and all variants (both recorded and new) simplifies notation and enables development of a multi-level classification model as described in the following.

	\noindent The multi-level multinomial logistic regression model is defined as follows:
	\begin{align*} \label{model:hier_multinom_init}
	P(c_i = k) &= \frac{\exp \left[\alpha_k + \sum_{j = 1}^d x_{i j} \betab_{j, k} \right]}{\sum_{k' = 1}^K \exp \left[\alpha_{k'} + \sum_{j = 1}^d x_{i j} \betab_{j, k'} \right]}, \text{ independently for } i = 1, \dots, n; \ k = 1, \dots, K \\
	\beta_{j, k} &\sim N(\ub_{j}^T \omegab_{\bullet, k}, \tau_j^2), \text{ independently for } j = 1, \dots, d; \ k = 1, \dots, K \\
	\omega_{l, k} &\sim N(0, \xi_l^2), \text{ independently for }  l = 1, \dots, p; \ k = 1, \dots, K  \numbereqn
	\end{align*}
	where  $\alphab = (\alpha_1, \dots, \alpha_K)^T \in \R^{K \times 1}$ is the vector of intercepts;  $\betab = (\beta_{j, k}) \in \R^{d \times K}$ is the matrix of individual variant regression coefficients with $k$-th column being $\betab_{\bullet, k} = (\beta_{1, k}, \dots, \beta_{d, k})^T \in \R^{d \times 1}$, $k = 1, \dots, K$ and the $j$-th row being $\betab_{j, \bullet} = (\beta_{j, 1}, \dots, \beta_{j, K}) \in \R^{1 \times K}$, $j = 1, \dots, d$; $\omegab = (\omega_{l, k}) \in \R^{p \times K}$ is the matrix of meta-regression coefficients, listing the coefficients associated with each meta-feature for various cancer types, with the $l$-th row of $\omegab$ being $\omegab_{l, \bullet}$, $l = 1, \dots, p$, and the $k$-th column being $\omegab_{\bullet, k}$, $k = 1, \dots, K$.

	The second level in \eqref{model:hier_multinom_init} encapsulates the key meta-feature regression step in our multi-level model. This step aids estimation of the meta-feature effects $\omegab_{\bullet, k}$ through the recorded variants. These estimated meta-feature effects $\hat \omegab_{\bullet, k}$ can subsequently ``predict'' the regression effects $\{\beta_{j, k}\}$ for a new variant $j$ ($j > d_1$) as $\hat \beta_{j, k} = \ub_j^T \hat \omegab_{\bullet, k}$. Note that because we only have access to the first $m$ tumors and the first $d_1$ variants in the training data, the regression effects $\{\beta_{d_1+1, k}, \dots, \beta_{d, k}\}$ for new variants are not identifiable and hence cannot be estimated in the training data set (only  $\{\beta_{1, k}, \dots, \beta_{d_1, k}\}$ can be estimated from the training data). A prediction of the cancer label $c_i$ for a test set individual $i$ ($i = m+1, \dots, n$) is made using  $\beta_{j, k}$'s estimated from the training data for $j \leq d_1$, and the $\beta_{j, k}$'s predicted via the meta-feature regression for $j > d_1$.
	
	Furthermore, this meta-feature regression step also aids a substantial dimension reduction of the predictor space. This becomes clearer in an alternative mixed-effect representation of the model described in the following section. This mixed-effect representation, in addition to aiding dimension reduction, also provides interpretation of the model parameters and aids scalable implementation.

	\subsection{Alternative mixed-effect representation}	 \label{sec:model_glmer}
	Instead of estimating the model parameters in \eqref{model:hier_multinom_init} directly, we first rewrite the model using  a mixed-effects representation of $\betab$, and then note how this alternative representation substantially simplifies implementation. Note that the second level in \eqref{model:hier_multinom_init} can be  expressed as 
	\begin{equation} \label{beta_representation}
	\beta_{j, k} = \beta_{j, k}^0 + \ub_j^T \omegab_{\bullet, k}
	\end{equation}
	where $\beta_{j, k}^0 \sim N(0, \tau_j^2)$. Writing $\sum_{j = 1}^d x_{i j} \beta^0_{j, k} = \x_i^T \betab_{\bullet, k}^0$, where $\betab_{\bullet, k}^0 = (\betab_{1, k}^0, \dots, \betab_{d, k}^0)$ and $\sum_{j = 1}^d x_{i j} \ub_j = \x_i^T \Ub$, we obtain the following alternative representation of \eqref{model:hier_multinom_init}:
	\begin{align*} \label{model:hier_multinom_glmer}
		P(c_i = k) &= \frac{\exp \left[\alpha_k + \x_i^T \betab^0_{\bullet, k} +  \x_i^T \Ub \omegab_{\bullet, k}\right]}{\sum_{k' = 1}^K \exp \left[\alpha_{k'} + \x_i^T \betab^0_{\bullet, k'} +  \x_i^T \Ub \omegab_{\bullet, k'} \right]},  i = 1, \dots, n; \ k = 1, \dots, K \\
		\beta^0_{j, k} &\sim N(0, \tau_j^2),  j = 1, \dots, d; \ k = 1, \dots, K \\
		\omega_{l, k} &\sim N(0, \xi_l^2),  l = 1, \dots, p; \ k = 1, \dots, K.  \numbereqn
	\end{align*}
	
	\noindent From representation \eqref{model:hier_multinom_glmer}, it follows that the penalized likelihood/un-normalized (i.e., without the normalizing constant) posterior density of $\alphab$, $\betab^0 = (\beta^0_{j, k})$ and $\omegab$ is essentially that of a regularized multinomial logistic regression likelihood with intercept $\alphab$, predictors $(\x_i^T, \x_i^T \Ub)$, associated regression coefficients $\betab^0$ and $\omegab$, and with ridge penalties  on $\betab^0$ and $\omegab$. For a fixed $j$ ($l$), the assumption of equal variance of $\beta^0_{j, k}$ ($\omega_{l, k})$ for all $k$ aids shrinkage/partial pooling in the estimation of the associated residual-regression (meta-regression) parameters, as is typical in multilevel models.
	
	The following observations are made as consequences of the reparameterization \eqref{model:hier_multinom_glmer}. First, through $\x_i^T\Ub$, the meta-features are now directly  ``connected'' to the classification (first level of the model). Note that, for a meta-feature $l$, $\x_i^T \Ub_{\bullet, l}$ quantifies the total mutational information or the total mutation burden in tumor $i$ that is described by that meta-feature, where $\Ub_{\bullet, l}$ denotes the $l$-th column of the meta-design matrix $\Ub$.  For example, if $\Ub_{\bullet, l}$ corresponds to the dummy variable indicating whether or not the variants are mutations at a specific gene $G_l$, (i.e., $u_{j, l} = 1$ if the $j$-th variant is indeed mutation of gene $G_l$, and 0 otherwise) then $\x_i^T \Ub_{\bullet, l}$ counts \emph{the total number of mutations at gene $G_l$ observed in tumor $i$}.  Consequently, the meta-regression parameter $\omega_{l, k}$ measures the effect of gene $G_l$ in describing the cancer type $k$ of a tumor.  Note that this ``connection'' naturally combines the effects of the individual variants and those of the genes where a mutation occurred, thus providing a refinement of the existing gene-based classification approach. Second, owing to the decomposition \eqref{beta_representation}, $\beta_{j, k}^0$ can be viewed as the $j$-th variant specific \emph{residual effect} that is not explained by the meta-regression.  As is the case in any mixed effects model, these  variant specific residual effects can only be measured for a variant that is present in the training data set, i.e., for $j \leq d_1$. (Indeed it can only be realistically utilized for the relatively few variants that occur somewhat frequently.) For $j > d_1$, the predicted residual effects $\beta_{j, k}^0$ are all zero. Third, observe that associated with each tumor $i$,  $\x_i^T\Ub$ essentially combines the presence/absence information of all training set/recorded variants through the meta-features; the condensed information is then used in classification. In other words, the product design-meta-design matrix $\X \Ub$ obtained by stacking the rows $\x_i^T \Ub$,  $i = 1, \dots, m$, aids a natural and potentially highly informative lower dimensional projection of the raw variant design matrix $\X$. Once information from all variants are properly condensed in the product design-meta-design matrix $\X \Ub$, the less informative variants can then be screened out from the design matrix $\X$ using an appropriate feature screening technique, such as an information-theory based feature screening method \citep{brown:2012} and/or sure independence screening \citep{fan:2008, fan:2017}.

	\subsection{Group Lasso Formulation for Parameter Estimation } \label{sec:estimation}
	Note that the primary parameters of interest in model \eqref{model:hier_multinom_glmer} (and \eqref{model:hier_multinom_init}) are $\alphab$, $\betab^0$, and $\omegab$ (and hence $\betab$), and the remaining are nuisance hyper-parameters. We propose estimation of the primary parameters through their marginal posterior modes (i.e., marginal a posteriori estimates). To this end, we include one additional hierarchical layer specifying the \emph{same gamma} prior distributions on $\{\tau_j^2\}$ and $\{\xi_l^2\}$, and then marginalize out $\{\tau_j^2\}$ and $\{\xi_l^2\}$ from the posterior  density to obtain a group-lasso  \citep{kyung:2010, yuan:lin:2006} regularized multinomial logistic likelihood for $\alphab$, $\betab^0$ and $\omegab$. In particular, we consider
	\[
	\tau_j^2, \xi_l^2 \sim  \text{independent Gamma} ((K+1)/2, \lambda^2/2), \quad j = 1, \dots, d, \ l = 1, \dots, p
	\]
	Then marginalizing out  $\{\tau_j^2\}$ and $\{\xi_l^2\}$ from the joint posterior density yields the following  group-lasso regularized multinomial logistic log-likelihood for $\alphab$, $\betab^0$ and $\omegab$:
	\begin{align*}  \label{likelihood_grouplasso}
		& \log \pi(\alphab, \betab^0, \omegab \mid \lambda, c_1, \dots, c_n, \x_1, \dots, \x_n) \\
	 	&= \sum_{i = 1}^n \sum_{k = 1}^K \one(c_i = k) \log \left( \frac{\exp \left[\alpha_k + \x_i^T \betab^0_{\bullet, k} +  \x_i^T \Ub \omegab_{\bullet, k}\right]}{\sum_{k' = 1}^K \exp \left[\alpha_{k'} + \x_i^T \betab^0_{\bullet, k'} +  \x_i^T \Ub \omegab_{\bullet, k'} \right]} \right) \\
		&\qquad \qquad  - \lambda \sum_{j = 1}^d \| \betab_{j, \bullet}^0 \|_2 - \lambda \sum_{l = 1}^p \| \omegab_{l, \bullet} \|_2 \numbereqn
	\end{align*} 
    where $\|\cdot\|_2$ denotes the squared norm, defined for $\bm v = (v_1, \dots, v_K)$ as $\|\bm v\|_2 = (\sum_{k = 1}^K v_k^2)^{1/2}$. Maximizing \eqref{likelihood_grouplasso} yields the group-lasso maximum marginal a posteriori estimates of $\alphab$, $\betab^0$ and $\omegab$ (and hence of $\betab$). An important  consequence of assigning a group lasso penalty is that for any $j$ (or any $l$), the maximum marginal a posteriori estimates of $\{\beta_{j, 1}^0, \dots, \beta_{j, K}^0\}$ (or $\{\omega_{l, 1}, \dots, \omega_{l, K}\})$ are either all exactly zero or all non-zero. In other words, the model aids variable selection at a collective variant and/or meta-feature level, i.e., either all cancer site-specific effects associated with a single variant and/or a single meta-feature are retained in the fitted model, or they are all dropped.

    \subsection{Prediction of Cancer Sites for New Tumors} \label{sec:prediction}
  	    Given the estimated residual regression parameters  $\hat \betab_{j, \bullet}^0$, $j = 1, \dots, d_1$, intercepts $\hat \alphab$, and  meta-regression parameters $\hat \omegab_{l, \bullet}$, $l = 1, \dots, p$, and the test data meta-design matrix $\Ub$, the cancer site prediction (classification) of a test set tumor $i$ with variant indicators $\x_i = (x_{i 1}, \dots, x_{i d})$ is performed as follows. First we set $\hat \betab_{j, \bullet}^0 = 0$ for all $j = d_1+1, \dots, d$, and obtain the full residual regression coefficient matrix $\hat \betab^0 = ((\hat \beta^0_{j, k}: j = 1, \dots, d, k = 1, \dots, K))$. Then for each $k$, we compute $\hat \eta_{i, k} = \hat \alpha_k + \x_i^T \hat \betab^0_{\bullet, k} + \x_i^T \Ub \hat \omegab_{\bullet, k}$, and subsequently normalize $\exp(\hat \eta_{i, k})$ to obtain the predicted probability of the tumor $i$ being of cancer type $k$: $\hat P(c_i = k) = \exp(\hat \eta_{i, k}) / \sum_{k' = 1}^K \exp(\hat \eta_{i, k'})$. These probabilities provide a soft classification for tumor $i$; a hard classification can be obtained by assigning tumor $i$ to class $k^* = \argmax_{k = 1, \dots, K} \hat P(c_i = k)$. It is of note that since the residual effects $\{\beta^0_{j, k}\}$ are all zero for a new variant $j \geq d_1 + 1$, the contribution of such a variant in $\hat \eta_{i, k}$ comes only through the $j$-th term of the sum $\x_i^T \Ub \hat \omegab_{\bullet, k}$, namely, $x_{ij} \ub_j^T \hat \omegab_{\bullet, k}$, which is the ``predicted'' (through the meta-regression) effect of the variant $j$, multiplied by the binary indicator $x_{ij}$ of that new variant being present in the $i$-th individual. That is, given a new variant in a new tumor, we first \emph{predict} the effect of that new variant based on its meta features 
  	      and then use the predicted effect in classification.

    \begin{remark}\label{remark:all_model_predictors}
      From representation \eqref{model:hier_multinom_glmer} it follows that our model essentially builds a classifier with two sets of predictors: the variant indicators $\x_i$ and the total mutation burden described by the meta-features $\x_i^TU$ (e.g., total number of mutations at each gene, and the total number of mutations associated with each single nucleotide change) in tumor $i$.
    \end{remark}
    	
    \begin{remark} \label{remark:identifiability}
		 It is of note that the symmetric representation of the multinomial logistic model provided in the first level of \eqref{model:hier_multinom_glmer} (or \eqref{model:hier_multinom_init}) does not identify the intercept and regression parameters by itself. For any values of the parameters $\{\alpha_k, \betab^0_{\bullet, k}, \omegab_{\bullet, k}\}_{k = 1, \dots, K}$, $\{\alpha_k-\gamma, \betab^0_{\bullet, k}-\gamma, \omegab_{\bullet, k} - \gamma\}_{k = 1, \dots, K}$ provide the same multinomial probabilities for any $\gamma \in (-\infty, \infty)$. However, the (prior) distributions specified in the second and third levels of \eqref{model:hier_multinom_init} identify these parameters, and hence aid their estimability. See, e.g., \citet{zhu:hastie:2004} and \citet{friedman:hastie:tibshirani:2010} for more details. 
	\end{remark}

	\begin{remark} \label{remark:interpretations}
		The symmetric representations of the multinomial logistic regression provided in the first levels of  \eqref{model:hier_multinom_glmer} and \eqref{model:hier_multinom_init} do not  aid direct interpretation in terms of the predictors in the model. 
		To achieve a log-odds ratio interpretation of the regression/meta-regression parameters, one first needs to define a reference category (say $1$). Then $\beta_{j, k} - \beta_{j, 1}$ (or $\omega_{j, k} - \omega_{j, 1}$)  measures the resulting change in the log odds of the cancer type being $k$ relative to being type $1$, when the $j$-th variant status is changed from absent to present, while keeping the statuses of all other variants fixed (or the total mutational burden captured through $l$-th meta-feature is changed by one unit, while keeping mutation burden described by other meta-features fixed). The residual effect parameters $\{\beta^0_{j, k}\}$ in \eqref{model:hier_multinom_glmer} are interpreted as effects of variant $j$ that are not explained by the meta-feature regression. Formally, $\beta^0_{j, k} - \beta^0_{j, 1}$ measures the unexplained/residual change  in the log odds of the cancer type being $k$ relative to being type $1$ after accounting for the ``expected''  change as ``predicted'' through the meta-feature regression  (viz., $\ub_j^T(\omegab_{\bullet, k} - \omegab_{\bullet, 1})$), when the $j$-th variant status is changed from being absent to being present, while keeping the statuses of all other variants fixed.    
	\end{remark}

	   \begin{remark} \label{remark:glmer_est}
	  It is to be noted that that the penalty parameter $\lambda$ in \eqref{likelihood_grouplasso} weights all  regression parameters equally. Hence, the predictors $\x_i$ and $\x_i^T\Ub$ should all be scaled prior to fitting the model \citep{tibshirani:1997}. For a given $\lambda$, the penalized likelihood \eqref{likelihood_grouplasso} can be efficiently optimized using a cyclical coordinate descent algorithm as described in \citet{friedman:2007pathwise} and \citet{friedman:hastie:tibshirani:2010}. This approach is highly scalable, and can  handle thousands of predictors (variants and meta features combined). The penalty parameter $\lambda$ can be chosen via a model-selection procedure such as cross-validation. The \texttt{R} package \texttt{glmnet} \citep{pkg_glmnet} provides an efficient implementation of this optimization approach, along with a cross-validation based selection of the penalty parameter. 
	\end{remark}

		\section{Data application: predicting cancer site of origin of TCGA tumors}
	\label{sec:application_tcga}
    
	We implemented our multi-level classification model using data from TCGA of the 7 highest prevalence tumor types, as determined by the SEER program of the National Cancer Institute \citep{duggan2016surveillance}.

    \subsection{Classification performance under cross-validation}  
    \label{sec:application_tcga_cv}
    
    We conducted five-fold cross-validation  to assess the prediction performance of our method. The entire set of 3702 tumors was partitioned into five random folds (maintaining the relative proportions of the seven cancer types as observed in the full data). Four out of the five folds were treated as the training set, and the fifth fold was treated as the test set in which the cancer site labels were masked, and all five combinations of test and training sets were considered for a given partition. The predicted class probabilities in the test sets were subsequently compared with the actual labels to evaluate prediction accuracy.
	
	To fit our model, we first obtained $x_{ij}$, the presence/absence indicator of variant $j$ in the $i$-th tumor in the training set, for all whole-exome variants that were observed at least once in the training data. (On a typical training set there were around 620,000 variants appearing at least once on average.) The meta-features involved dummy coded presence/absence indicators of the 584 clinically relevant cancer genes constituting the oncoKB list \citep{chakravarty:2017:oncokb} and the 96 single base nucleotide substitution signatures.  The single base nucleotide substitution signature is the type of alteration (e.g., C to T transitions) in the trinucleotide context. For each mutated locus, there are six classes of base substitution (C>A, C>G, C>T, T>A, T>C, T>G), together with the nucleotide base immediately 5' and 3' to each mutated base, yielding a total of 96 possible single base nucleotide substitution signature types. These alterations are the key elements in classical mutation signature analysis \citep{alexandrov:2013}, and reveal the mutagenic processes specific to certain cancer types (e.g., the C>T and C>G mutations at TpCpN trinucleotides are signature of APOBEC defect which is prevalent in bladder cancer). The design matrix $\X = ((x_{ij}))$ and the product design-meta-design matrix $\X \Ub$ were subsequently obtained. As described in Section~\ref{sec:model_glmer}, the $(i, l)$-th element of the product design-meta-design matrix $\X \Ub$ is the total burden in the $i$-th tumor that is associated with the $l$-th meta feature (a specific gene or corresponding to a specific neucleotide change signature), and thus can be computed without actually constructing the meta-design matrix $\Ub$ and performing the computationally expensive high dimensional matrix multiplication of $\X$ and $\Ub$.
    
    After obtaining $\X$ and $\X \Ub$, we performed a screening of the raw variants to find and retain the most discriminating ones in the sparse multilevel multinomial logistic model. For this, we used a normalized mutual information (NMI)  based feature screening \citep{strehl:ghosh:2003, cover:thomas:2012} which can be regarded as a generalization of the feature importance based sure independence screening approach described in \citet{fan:2008}. See Web Appendix A for the definition and a  brief note on normalized mutual information. We estimated the NMI with cancer types separately for all variants in the training data, and filtered out from the design matrix $\X$ all variants with NMI ranking $\geq 250$.

	Along with our multi-level model with observed variants as predictors, and the associated gene labels and mutation footprints as meta-features, we also considered the following three competing models to compare their relative  classification performances: (a)  a group-lasso penalized sparse multinomial logistic classifier with the predictors being indicators of the presence/absence of mutations at each of the 584 oncoKB genes,  (b)  a random forest gene level classifier  with the same predictors as in (a), and  (c) a recorded-variant-only sparse multinomial logistic classifier with group lasso penalty. Thus (a) tells us the achievable accuracy, based on a generalized linear model, using only gene-level information; (b) is a non-parametric version of (a) with the generalized linear and parametric assumptions removed (and thus has a potentially better predictive ability than (a)), and (c) tells us  the extent of diagnostic accuracy using only recorded variant specific information. For (c), we first performed a similar feature screening as done in our multilevel model, but kept all variants with NMI ranking $< 1000$. All group lasso penalty parameters were chosen via separate (independent) 10 fold cross-validations. All computations were done in statistical software \texttt{R} v 3.4.4 \citep{R_software}, and the packages \texttt{tidyverse} \citep{pkg_tidyverse}, \texttt{data.table} \citep{pkg_datatable}, \texttt{Matrix} \citep{pkg_Matrix}, \texttt{glmnet} \citep{pkg_glmnet}, and \texttt{precrec} \citep{pkg_precrec}  were used. 
	
	After fitting the four models on each training data set, the predicted probabilities of being classified into each of the seven cancer types were computed for each tumor in the associated test set for each method. These predicted probabilities were subsequently combined from all five folds. We repeated the entire cross-validation experiment (with all four models) on 100 different random partitions (folds) of the tumors, and then collected from each cross-validation the classification probabilities from all four models, and in each of the seven cancer categories. 
	
	To compare predictive performances of the four competing models we considered the resulting one-vs-rest comparison in each cancer category based on the classification probabilities. On each cross-validation, for each method, in each one-vs-rest comparison,  we obtained a precision-recall curve by computing the precision (also called positive predictive value) and recall (also sensitivity) associated with varying binary classification thresholds \citep{pkg_precrec}. It is of note that precision-recall curves are substantially more informative than receiver-operator characteristic (ROC) curves in a classification problem with imbalanced classes, which is the case here due to the small sample size of the PAAD class (176, 4.76\% of 3702). The corresponding area under the curve (AUC) was computed as a summary of each one-vs-rest classification performance, and the average AUCs from all seven one-vs-rest classifications were subsequently obtained as  univariate measures of overall classification performance for each classifier. The means and the standard deviations of these AUCs (and the overall average AUC)  across all cross-validations were computed to quantify the variability of these measures across cross-validations. These values are shown in Table~\ref{table:aucs}. (The actual precision-recall curves are shown in Web Figure~1).

	\begin{table}[ht]
		\setlength{\tabcolsep}{8pt} 
		\renewcommand{\arraystretch}{1.25} 
		\centering
	
        \label{table:aucs}
        
		\begin{tabular}{|l|r|r|r|r|}
            \toprule
            \multicolumn{1}{|p{3cm}|}{\centering } 
            & \multicolumn{1}{|p{3cm}|}{\centering Seen+Unseen \\ variants SMML} 
            & \multicolumn{1}{|p{3cm}|}{\centering Gene \\ RF} 
            & \multicolumn{1}{|p{3cm}|}{\centering Gene \\ SML}
            & \multicolumn{1}{|p{3cm}|}{\centering Recorded \\ variants SML} \\
            \midrule
            BLCA & 0.72 (0.01) & 0.64 (0.011) & 0.62 (0.012) & 0.51 (0.0097) \\
            BRCA & 0.72 (0.0077) & 0.67 (0.0041) & 0.72 (0.0057) & 0.59 (0.0053) \\
            COADREAD & 0.81 (0.0078) & 0.73 (0.0038) & 0.75 (0.0061) & 0.70 (0.009) \\
            LUAD & 0.84 (0.0028) & 0.61 (0.0088) & 0.63 (0.0088) & 0.45 (0.0079) \\
            PAAD & 0.56 (0.014) & 0.54 (0.0073) & 0.52 (0.016) & 0.44 (0.013) \\
            PRAD & 0.53 (0.01) & 0.45 (0.01) & 0.51 (0.0075) & 0.26 (0.0071) \\
            SKCM & 0.94 (0.0039) & 0.79 (0.0036) & 0.84 (0.006) & 0.84 (0.0097) \\
            Average & 0.74 (0.006) & 0.63 (0.003) & 0.66 (0.0085) & 0.54 (0.0084) \\
            \bottomrule
        \end{tabular}
			\caption{Area under the precision-recall curve for each one-vs-rest classification (first seven rows) and their average (last row)  obtained from the classification probabilities assigned by each classifier. The numbers inside a cell (outside the parentheses) represents the cross-validation mean of the area under the curves (AUCs) for the associated one-vs-rest classification (first seven rows, labeled at the left border), or the cross-validation mean of the average AUC (from all one-vs-rest classification, last row) for the corresponding classifier (labeled at the top border), and the numbers inside the parentheses are the corresponding (cross-validation) standard deviations.}
	\end{table}
	
	For each of the four classifiers, a hard classification rule was obtained by assigning each tumor to the class with highest probability. The precision-recall values for these hard classifiers were computed under cross validations, and results are shown in Web Table~1 in Web Appendix~C. These results show a broadly similar pattern in terms of relative performances of the four methods. 
    
    From the cross validation results displayed in   Table~\ref{table:aucs}, and also Web Figure~1 and Web Table 1, it is clear that the multilevel model consistently, and often substantially, outperforms the other three classifiers in each one-vs-rest comparison as quantified through the AUCs of the precision-recall curves and the curves themselves, Web Figure~1. In terms of the improvement in AUC over its closest competitor, the largest gains are observed in the cancer categories LUAD, BLCA, SKCM and COADREAD; in the remaining three cancer types, PAAD, BRCA and PRAD, there are still noticeable positive gains, although not as dramatic as the former three. [None of four classifiers performs well in the cancer category PRAD, which is not unexpected, as PRAD is not a very mutation driven cancer.] This illustrates the importance of the mutation signatures as meta-features in our classification model (discussed in more detail in the next section). The gene-level classifiers (random forest and sparse multinomial logistic) have similar performances, while the recorded variant only model clearly has a noticeably inferior classification performance.

    \subsection{Obtaining covariate effects from a full data fit}    
    
    The cross-validation experiments illustrate the substantial gain in prediction accuracy achieved by the full multilevel model  consisting of the effects of individual variants, in addition to the 584 genes and 96 single base nucleotide substitution meta-features. To gain insights on the effects of these factors we refitted the (``full'') model on the whole TCGA data set with all 3702 tumors. As before, we first computed the product design-meta-design matrix $\X \Ub$ by counting the total number of variants corresponding to each meta-feature in a tumor. We then obtained the ranks of the NMI values with cancer types for all 811,497 variants observed in the data, and screened out all variants with NMI rank $\ge 250$. The group lasso penalty parameter was chosen via a 10-fold cross-validation (using \texttt{glmnet}). The penalized regression parameter estimates quantifying the variant, gene and single base nucleotide substitution signature effects were then obtained from the fitted model. To aid interpretation, we considered PRAD (prostate cancer) as the reference group, as this group has tumors with low mutation burden in general, and computed the log odds ratios of being classified into each of the other six cancer sites (see Remark~\ref{remark:interpretations}).  To aid comparison, each log odds ratio estimate was subsequently multiplied by the associated standard deviation of the predictor and then exponentiated, providing odds ratios for one standard deviation changes in the predictors. 
    
\begin{figure}
		\centering	
		\includegraphics[width=\linewidth]{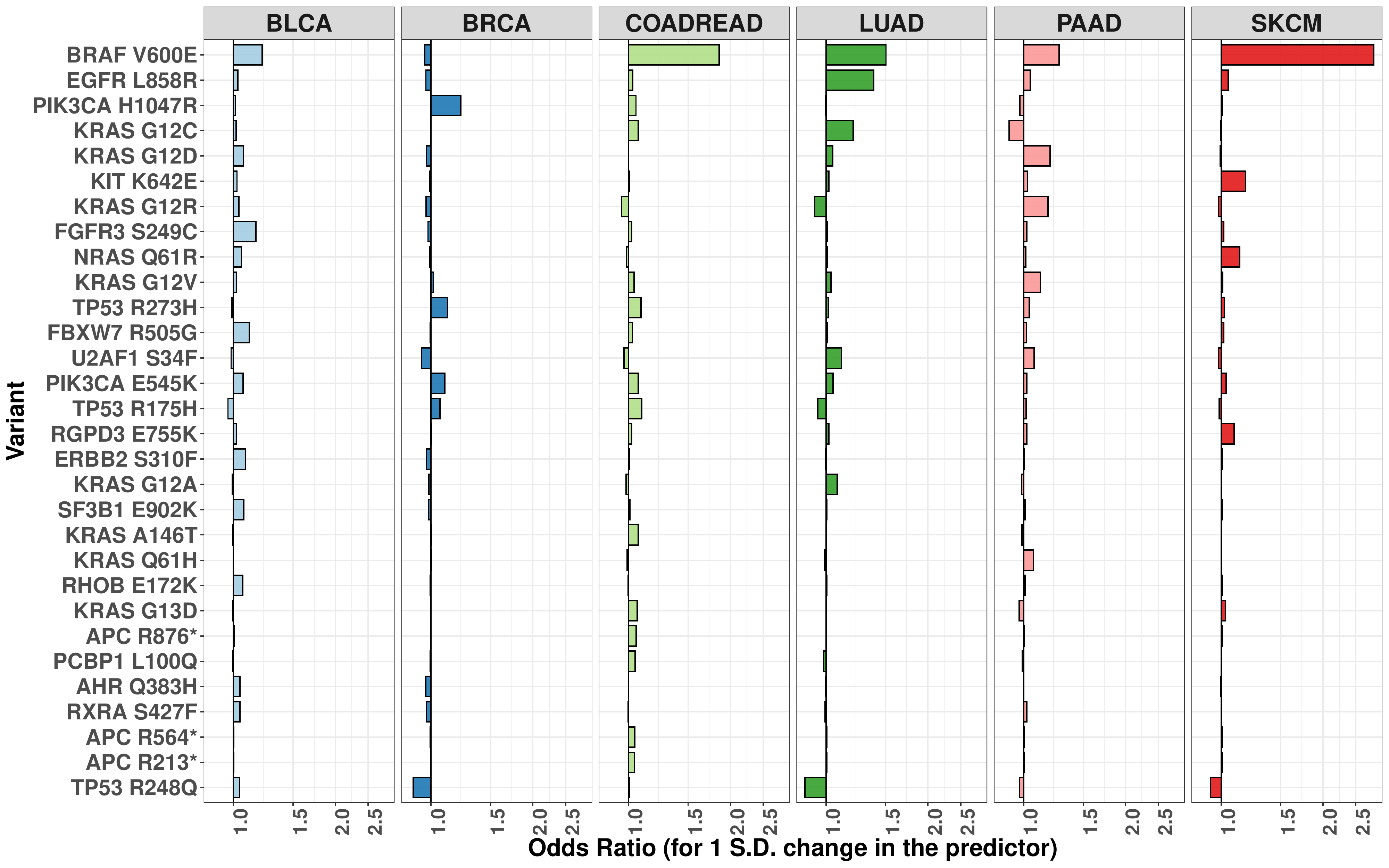}  
		\caption{Treating prostate cancer (PRAD) as the reference category, the (residual) ratio of odds of classifying a tumor into one of the other six cancer categories, for a one standard deviation change in the presence/absence of raw variants are plotted for 30 variants with largest (residual) odds ratios. Each horizontal bar indicates the magnitude of the estimated odds ratios (in a log scale) associated with the corresponding variant (row) and in each cancer type (column).}
        \label{fig:beta0}
	\end{figure}

    We first considered the ``residual'' effects $\betab_0$ of the raw variants included in the model (retained after normalized mutual information based screening), i.e., the variant specific effects that are not explained by the meta-features. Unsurprisingly, most of the estimated residual effects were zero/small, as a consequence of the group lasso penalty. With PRAD as the reference group, we computed the corresponding ratio of odds of being classified into one of the other six categories. In Figure~\ref{fig:beta0}, 30 variants with the largest estimated odds ratios are displayed.  It appears that the largest estimated odds ratio is observed in SKCM for \emph{BRAF} V600E, a mutation that has been linked to skin cancer in multiple studies \citep{Bailey:2018}. It is also of note that a  handful of \emph{KRAS} variants, such as G12C, G12D, G12R and G12V, have strong non-zero effects on cancer classification. These findings are in concordance with the existing scientific literature in that we see the G12C variant exhibiting its largest positive effects in lung adenocarcinoma (LUAD), while G12D, G12R and G12V display their largest effects on pancreatic cancer (PAAD). These variant specific effects cannot be captured in a gene-only model that identifies a mutation only through the associated gene. 
     
\begin{figure}
		\centering	
		\includegraphics[width=\textwidth]{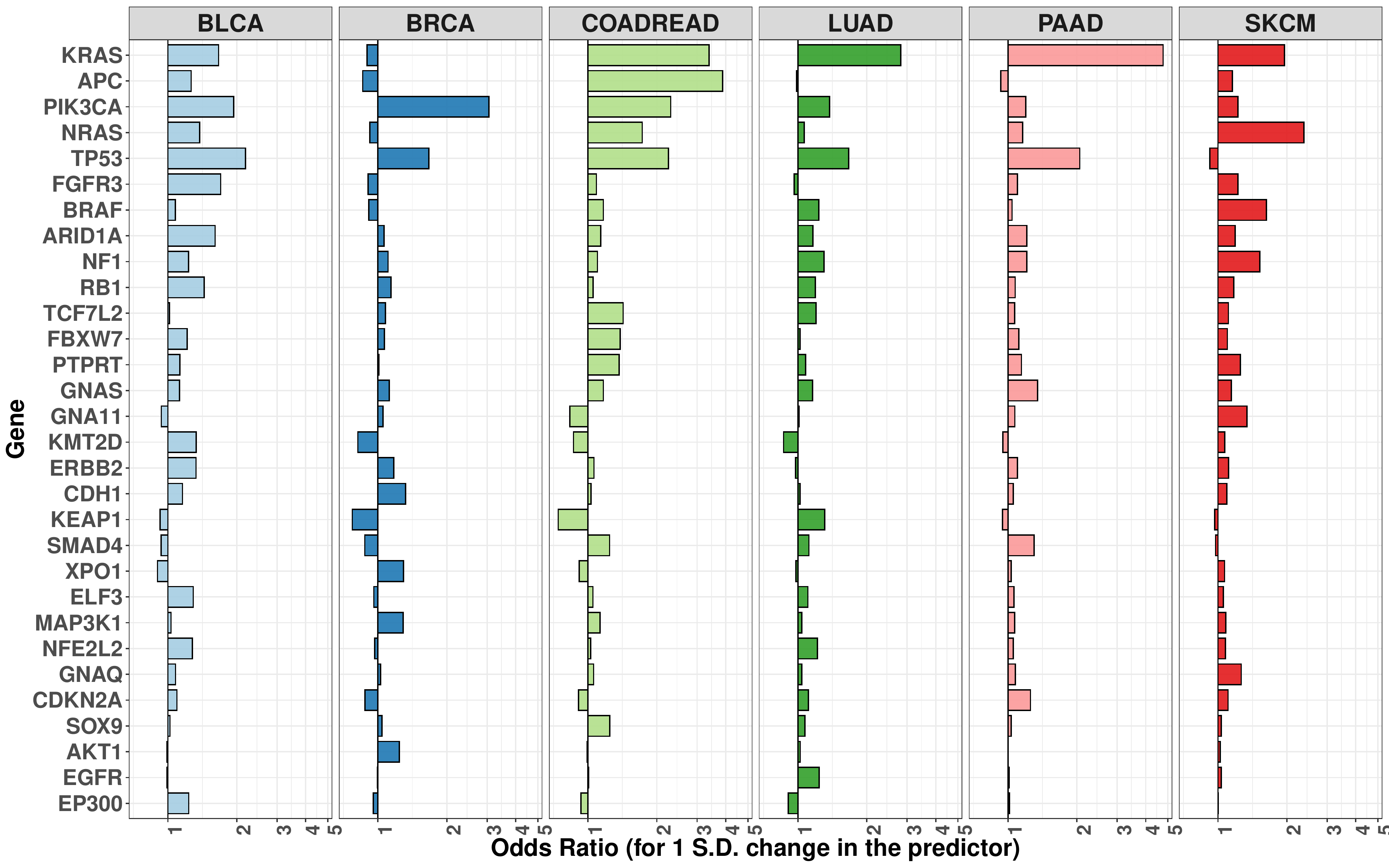}
        \caption{Treating prostate cancer (PRAD) as the reference category, the ratio of odds of classifying a tumor into one of the other six cancer categories, for a one standard deviation change in the corresponding level of gene are plotted for 30 genes with largest odds ratios. Each horizontal bar indicates the magnitude of the  odds ratios (in a log scale) for each of the 30 genes (row), and in each cancer type (column).}
        \label{fig:omega_gene}
	\end{figure}

	Next, to visually assess the relative effects of the genes as meta-features in our cancer classification model, we considered the 30 genes with the largest odds ratios of classifying a tumor into one of the six cancer categories other than PRAD (for one standard deviation changes in the gene levels), and plotted odds ratios in Figure~\ref{fig:omega_gene}. The figure shows large values for genes whose mutations are known to be highly tissue specific. For example, mutations at APC have been associated with colorectal cancer (CODEREAD), those at BRAF and NRAS with melanoma skin cancer (SKCM), and those at KRAS with both lung (LUAD) and pancreatic (PAAD) cancers. 
	
\begin{figure}
		\centering
		\includegraphics[width=\textwidth]{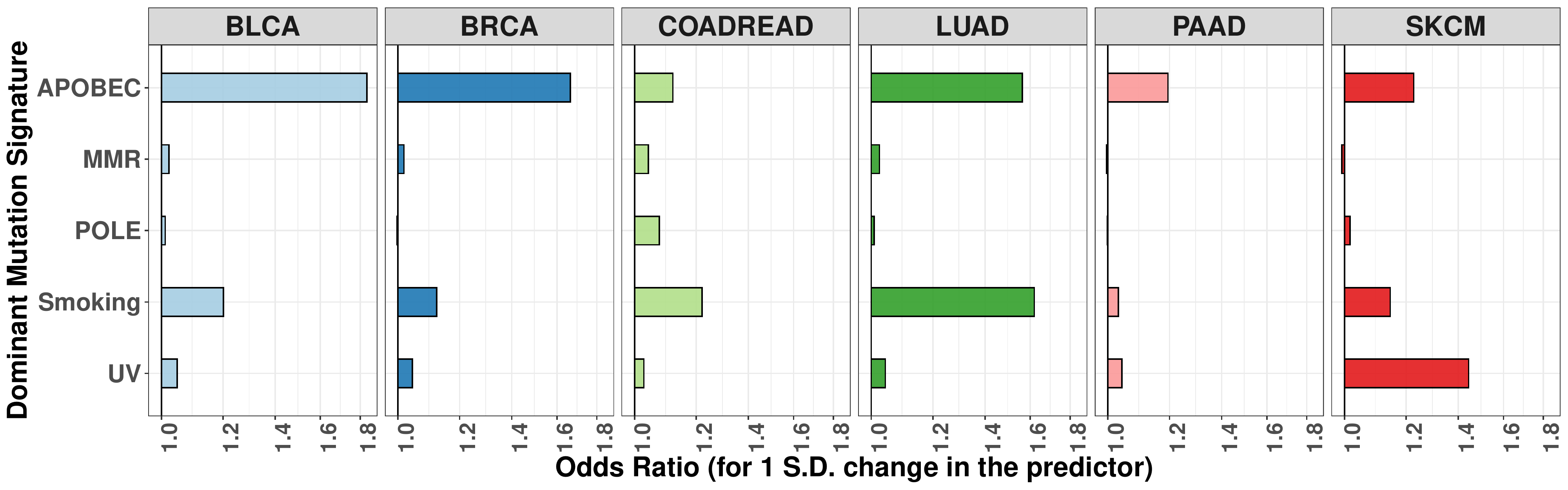}
        \caption{Treating prostate cancer (PRAD) as the reference category, odds ratios of classifying a tumor into one of the other six cancer categories, for a one standard deviation change in the corresponding levels of the 5 dominant mutation signature groups are plotted. Each horizontal bar indicates the magnitude of odds ratio (in a log scale) for each of the 5 dominant mutation signature groups (rows), and in each cancer type (column).}
		\label{fig:omega_mutsig}
	\end{figure}

	Finally, to visualize the relative effects of the single base nucleotide substitution  signatures in predicting cancer types, we obtained odds ratios for dominant mutation signature groups. We considered five dominant signature groups, viz., APOBEC, MMR, POLE, Smoking and UV, each being a weighted average of the 96 raw single base substitution signatures \citep{alexandrov:2013}, and computed the corresponding weighted average of the estimated meta-regression coefficients, $\tilde \omegab_h = {\bm v_h}^T \hat \omegab_{MS-1:96, \bullet}$, and the weighted design-meta-design matrix, $\X \Ub_{\bullet, MS-1:96} \bm v_h$, for each of these signature groups where $\bm v_h$ denotes the weights associated with the $h$-th dominant signature ($h \in$ \{APOBEC, MMR, POLE, Smoking, UV\}). Here  $\hat \omegab_{MS-1:96, \bullet}$ ($\X \Ub_{\bullet, MS-1:96}$) is a sub-matrix of the estimated meta-regression coefficient matrix $\hat \omegab$ (design-meta-design matrix $\X \Ub$), with rows (columns) corresponding to the 96 single-base substitution signatures. The weighted effects $\tilde \omegab_h$ were then remeasured for one standard deviation changes in these dominant mutation signatures, by multiplying with standard deviations of the columns of  $\X \Ub_{\bullet, MS-1:96}$. Finally, treating PRAD as the baseline category, the odds ratios for classifying a tumor into one of the other six categories were computed for one standard deviation changes in these dominant signature groups, and were plotted in Figure~\ref{fig:omega_mutsig}. Large odds ratios are obtained for the UV signature in melanoma skin cancer (SKCM), the smoking signature in lung cancer (LUAD), the MMR signatures in both colorectal cancer (COADREAD), and  the APOBEC signatures in bladder cancer (BLCA), breast cancer (BRCA), and lung cancer (LUAD). These findings are in perfect alignment with the existing scientific knowledge \citep{alexandrov:2013}.

	\section{Discussion} \label{sec:discuss}
	
	We considered a multi-level multinomial logistic regression model for predicting/classifying primary sites of origin of tumors with unknown primary. Our model incorporates information on hitherto unseen mutations through the associated genes and mutation footprints, treated as meta-features in the model, and uses this information in the classification. We observed substantial gains in classification accuracy over approaches that do not take advantage of these meta-features, such as a recorded variant only classifier, and a gene-level classifier. Along with the variants, we investigated the roles of the associated genes and the single base nucleotide substitution mutation footprints as predictors in our model, and gained further insights on the discriminating abilities of these attributes. 
	
    There are a number of future directions in which our current approach and findings could be extended. First, we limited the scope of our analysis by considering seven common cancer categories in the TCGA data for classification, using the whole-exome variants as predictors, and using only gene memberships and single base mutation signatures as meta-features. These were selected for expository purposes, since these tumor types had adequate sample sizes and well known mutational features. In a clinical study, interests may lie in a larger number of cancer types with more clinical relevance, and inclusion of a reference (no cancer) category may be of practical importance, e.g., for a ctDNA study where the patient may not have cancer at all. It may also be of interest to include extra-exome variants, obtained for example from a whole genome sequencing study, as predictors to assess their discriminating/predictive abilities and to see if they improve the performance. Finally, additional meta features for a variant, such as its replication time, exome size, expression levels etc. could potentially help extract further information from the rare and unseen variants, thereby bolstering predictive abilities of the model. Our approach could be easily extended to address these scientific problems. If the total number of cancer categories is sufficiently large, then a sparse group lasso penalty \citep{simon:2013}, instead of the group lasso penalty considered in this paper, could lead to a potentially better classification accuracy. Second, instead of the penalized likelihood based optimization approach (maximum a posteriori estimation), a full Bayesian approach could be adopted for estimation of model parameters. Then, instead of predicting the residual effect of a hitherto unseen variant to be exactly equal to zero, one would consider an ensemble of possible values via random generation from the associated posterior, as obtained from the meta-regression, and combine those with posterior samples for residual regression parameters associated with ``recorded'' variants, and those for the meta-regression parameters.  Note that to properly account for the sparsity in the predictor space in a full Bayesian framework one would need to use a sparse Bayesian prior e.g., the spike and slab prior or a global-local shrinkage prior, such as the horseshoe prior \citep{carvalho:2009}, instead of the lasso type priors considered in this study.  Finally, a non-parametric Bayesian approach to sparse multinomial logistic regression, as considered in \citet{burgette:reiter:2013},  might also be profitably combined with the hierarchical structure considered in this paper. This approach may lead to additional insights such as which mutations and/or meta-features are effectively ``similar'' in classifying tumors by appropriately clustering the mutations into a number of distinct groups.

	\bibliographystyle{biom}
	\bibliography{biblio}
	
	\section*{Supporting Information}
	Web Appendices, Tables, Figures, and a zip file containing the data and the \texttt{R} scripts used for the analyses are available in the Supporting Information section at the end of the article.\vspace*{-8pt}
	
	\section*{Data Availability Statement}
	The TCGA data that support the findings of this study are openly available in the GDC Data Portal of National Cancer Institute at https://portal.gdc.cancer.gov/.

	
	
	
	
	
	\label{lastpage}
	
\end{document}